\documentclass[reprint,prl,superscriptaddress]{revtex4-1}

\pdfoutput=1
\usepackage{graphicx}
\usepackage{dcolumn}
\usepackage{amsmath}
\usepackage{CJK}
\renewcommand{\vec}[1]{\mathbf{#1}}
\newcommand{\rr}{\vec{R}}

\begin{document}

\begin{CJK*}{GBK}{}

\title{A first-principles approach to nonlinear lattice dynamics: Anomalous spectra in PbTe}

\author{Yue Chen }
%\author{Yue Chen (³ÂÔÁ)}
\email[Corresponding author: ]{yuechen@columbia.edu}
\affiliation{Department of Applied Physics and Applied Mathematics, Columbia University, New York, NY 10027, USA}
\author{Xinyuan Ai }
%\author{Xinyuan Ai (°¬Ü°Ôª)}
\affiliation{Department of Physics, Columbia University, New York, NY 10027, USA}
\author{C. A. Marianetti}
\affiliation{Department of Applied Physics and Applied Mathematics, Columbia University, New York, NY 10027, USA}

\date{\today}

\begin{abstract}
Here we introduce a new approach to compute the finite temperature lattice dynamics from first-principles via the newly developed slave mode expansion. We
study PbTe where inelastic neutron scattering (INS) reveals strong signatures of
nonlinearity as evidenced by  anomalous features which emerge in the phonon
spectra at finite temperature.  Using our slave mode expansion in the classical limit, we compute the
vibrational spectra and show remarkable agreement with temperature dependent
INS measurements.  Furthermore, we resolve experimental controversy by showing
that there are no appreciable local nor global spontaneously broken symmetries at finite temperature
and that the anomalous spectral features simply arise from two anharmonic interactions.
Our approach should be broadly applicable across the periodic table.
\end{abstract}

\maketitle

\end{CJK*}
Inelastic neutron scattering (INS) is a fundamental probe of materials which
has allowed a unique view into the most basic aspects of mechanical behavior \cite{Lovesey}.
With the advent of the Spallation Neutron Source, massive amounts of data can be
accumulated at an unprecedented rate,
allowing for an extremely detailed inspection of phenomena throughout reciprocal space. Despite the enormous success of this experimental method,
theory has greatly lagged behind in the context of vibrations. At present, one cannot compute the
temperature dependence of the phonon spectrum for materials with
appreciable phonon interactions from first-principles. Moreover, this
scenario cannot even be routinely handled in the  classical limit
for indirect reasons.
This assertion is empirically well illustrated by the strongly interacting phonon material
PbTe. INS measurements demonstrated signatures of strong
interactions in the temperature dependence of the phonon spectra \cite{Delaire2011,Jensen2012}, attracting
significant attention to this system, yet complementary theoretical predictions have not yet been made.
In this paper, we circumvent previous theoretical limitations, and resolve the experimental
anomalies in PbTe.

In many materials, density functional
theory (DFT) is expected to describe the structural energetics to a high
degree of accuracy. Therefore, the interatomic potential generated via solving
the DFT equations (assuming the Born-Oppenheimer approximation) should be
reliable to perform both quantum and classical dynamics of the nuclei. However,
the problem is that the scaling of DFT is sufficiently prohibitive to forbid
this in many scenarios. Even at the level of classical mechanics (aka.
\emph{ab initio} molecular dynamics), one would need a very large unit cell along with many time steps. In most cases
this prevents one from directly being able to compute even the classical vibrational
spectrum at finite temperature, as evidenced by the sparse number of such publications in the literature. There are a variety
of approaches that attempt to circumvent this problem. One approach is the use of an empirical potential in place of
DFT. If an accurate empirical potential exists this is acceptable, but this is generally not the case.
Another approach would be in the spirit of static mean-field theories based on DFT energetics.
Two such approaches are the Self-Consistent \emph{Ab Initio} Lattice Dynamics (SCAILD) method \cite{Souvatzis2009} and the approach of Hellman \emph{et al.} \cite{Hellman2011},
both of which can capture the static phonon renormalization as a function of temperature.
However,
these approaches are seriously limited in that they are static approaches which cannot account for phonon lifetimes, let alone
strongly incoherent behavior. The well known quasiharmonic approximation has the same limitations as the former, while often being
less reliable \cite{Bozin2010}. Another avenue is using more efficient electronic structure methods. Tight-binding based approaches \cite{Papaconstantopoulos2003413} typically have excessive errors even at the level of phonon frequencies.
Linear-scaling DFT approaches \cite{Bowler2012036503,Vandevondele20123565} have definitely bridged the length scale, but the timescale is still a formidable
barrier to obtaining a robust spectra, in addition to the open questions regarding precision of the basis set.
A final approach would be to Taylor series expand the DFT energies as a function of the atomic
displacements (see Ref.~\cite{Esfarjani2008} for a detailed overview). This would retain all the accuracy of DFT to a certain range and order, while having a relatively negligible cost:
\begin{equation}
V= V_H + \sum_{ijk}\Psi_{ijk}u_{i}u_{j}u_{k}
+ \sum_{ijkl}\chi_{ijkl}u_{i}u_{j}u_{k}u_{l} + \cdots \label{eq.energyExp}
\end{equation}
where $V_H$ is the harmonic contribution; $\Psi$ and $\chi$ denote the 3$^{rd}$ and 4$^{th}$ order force constants, respectively; $u$ is the atomic displacement,
while index $i$, $j$, $k$ and $l$ label both the atom and the cartesian coordinate.
There is a long history of performing such an expansion in various degrees of sophistication, and this approach and the
details of all the nontrivial symmetries that must be satisfied have been well
outlined \cite{Esfarjani2008}.  However, the very few demonstrably robust
executions of this straightforward Taylor series highlight the difficulties of
computing many expansion coefficients while simultaneously satisfying all the
symmetries. A Taylor series to third order and within nearest neighbor was used
in PbTe \cite{Shiga2012}, and while this did reproduce the thermal conductivity,
the spectra was not computed. Recently, we have introduced a new approach called the Slave Mode
Expansion \cite{Ai2014}, which essentially makes the Taylor series substantially more
tractable as evidenced by our ability to straightforwardly compute all 358
anharmonic expansion coefficients which exist within an octahedra (i.e. up to next-nearest neighbor interactions) up to fourth
order. In this paper, we demonstrate the utility of our approach by computing
the imaginary part of the phonon Green's function in the classical limit for PbTe,
resolving the anomalous behavior in PbTe and ending experimental discrepancy.

PbTe has garnered significant interest as a thermoelectric material \cite{Heremans2008554}. Bate \emph{et al.} \cite{Bate1970} found that the undistorted ground state structure of PbTe
transforms to a paraelectric phase at elevated temperatures, which contradicts
the conventional picture of ferroelectric-paraelectric transformation. More recently,
Bo\v{z}in \emph{et al.} \cite{Bozin2010} proposed that the
emergence of stable local dipoles from the undistorted structure in PbTe based on the analysis of the pair distribution function (PDF). In addition, the phonon dispersions in PbTe were recently measured using INS at different temperatures \cite{Delaire2011,Jensen2012}. It was found that the phonon scattering in PbTe
is strongly anharmonic and displays anomalous behavior at elevated
temperatures, i.e. the `waterfall' effect of the transverse optical (TO) phonon
mode at $\Gamma$ point and the avoided-crossing behavior between the longitude
acoustic (LA) and TO branches in the $\Gamma \rightarrow \textrm{X}$ direction.
More importantly, a new spectral feature emerges at the zone center at finite temperature,
which is a clear signature of strong interactions and requires a theoretical explanation.

Theory has already made a number of contributions in understanding PbTe.
The phonon dispersions have been studied in numerous publications by using either the finite displacement \cite{Shiga2012,Zhang2009} or linear response methods \cite{An2008,Romero2008,Tian2012} based on DFT. The SCAILD method successfully captured the qualitative trend for the shift of the TO mode spectra to higher
frequency with increasing
temperature \cite{Bozin2010}, while the quasi-harmonic approximation failed
by predicting the opposite \cite{An2008}. However, SCAILD is incapable
of predicting a redistribution of spectral weight and therefore it cannot capture the emergence
of the anomalous spectra at the zone center (i.e. it only shifts the the delta function without any broadening).
\emph{Ab initio} molecular dynamics
(AIMD) simulations reproduced the
non-Gaussian asymmetric behaviors of the PDF peaks in PbTe \cite{Zhang2011}, however, due to the
limited number of atoms and short-time scale that AIMD can currently simulate,
the vibrational spectra can not be effectively evaluated from the trajectories. In
addition, empirical interatomic potentials, which have been developed for PbTe \cite{Qiu2012,Chonan2006}, fail to capture the most important features of the
phonon dispersions, i.e. an extremely soft TO mode near the $\Gamma$ point.
Therefore, none of the existing techniques can handle strongly incoherent behavior
demonstrated in PbTe, where
at certain wavevectors the concept
of a mode lifetime is rendered meaningless.

In order to compute the vibrational spectra from DFT forces, we have used the
Slave Mode Expansion \cite{Ai2014} to construct a polynomial interatomic
potential containing all terms within next-nearest neighbors and up to fourth
order for PbTe.  We show that there are 56 terms at 3$^{rd}$ order and 302 at
4$^{th}$ order, and this is dictated purely by symmetry. Density functional
theory has been used to compute all 358 terms, and we have demonstrated that
these have a high fidelity and that longer range interactions are small \cite{Ai2014}. Moreover, we have proven that one can
perform a unitary transformation to construct a \emph{minimal} model of
anharmonicity which captures the vast majority of the physics with simply one
term at 3$^{rd}$ order and one term at 4$^{th}$ order. The resulting minimal
potential can be compactly expressed as follows:
\begin{align}\label{potential} \nonumber
&V=V_H+ \\ \nonumber
&\Phi_{3}\sum_{\rr }
(\phi_{\rr x_-}^3 - \phi_{\rr x_+}^3 + \phi_{\rr y_-}^3 - \phi_{\rr y_+}^3 + \phi_{\rr z_-}^3 - \phi_{\rr z_+}^3) + \\
& \Phi_{4}\sum_{\rr }
(\phi_{\rr x_-}^4 + \phi_{\rr x_+}^4 + \phi_{\rr y_-}^4 + \phi_{\rr y_+}^4 + \phi_{\rr z_-}^4 + \phi_{\rr z_+}^4)  \\ \nonumber
& \textrm{and:} \\ \nonumber
& \begin{array}{lr}
\phi_{\rr z_-}=\frac{1}{\sqrt{2}}(u^{\rr+\bf{a_1}}_{Te,z}      -u^\rr_{Pb,z}) &
\phi_{\rr z_+}=\frac{1}{\sqrt{2}}(u^{\rr+\bf{a_2+a_3}}_{Te,z}  -u^\rr_{Pb,z}) \\
\phi_{\rr x_-}=\frac{1}{\sqrt{2}}(u^{\rr+\bf{a_2}}_{Te,x}      -u^\rr_{Pb,x}) &
\phi_{\rr x_+}=\frac{1}{\sqrt{2}}(u^{\rr+\bf{a_1+a_3}}_{Te,x}  -u^\rr_{Pb,x}) \\
\phi_{\rr y_-}=\frac{1}{\sqrt{2}}(u^{\rr+\bf{a_3}}_{Te,y}      -u^\rr_{Pb,y}) &
\phi_{\rr y_+}=\frac{1}{\sqrt{2}}(u^{\rr+\bf{a_1+a_2}}_{Te,y}  -u^\rr_{Pb,y})
\end{array}
\end{align}
Where $\phi$ are the slave modes which
are simply dimers in this case, and $\bf{a_i}$
are the primitive lattice vectors of PbTe: ${\bf a_1}=a/2(1,1,0)$, ${\bf a_2}=a/2(0,1,1)$, and ${\bf a_3}=a/2(1,0,1)$.
There are six dimer slave modes per primitive unit cell, one corresponding to each Pb-Te octahedral bond,
and these are simply a displacement difference between corresponding vectors of Pb and Te.
The
values for the expansion coefficients are found to be
$\Phi_{3}=2.68$ eV/{\AA}$^{3}$ and $\Phi_{4}=3.70$ eV/{\AA}$^{4}$, respectively. These parameters were computed without spin-orbit coupling, and this is known to overpredict the $T=0$ K TO-mode phonon frequency \cite{Romero2008}, which should be accounted for when comparing to experiment.

Classical dynamics simulations are performed using Eq.~\ref{potential}.
A $10\times10\times10$
supercell of the conventional unit cell containing 8000 atoms is used in the simulations with periodic
boundary conditions. The microcanonical ensemble and a time step of 5 fs are
applied. The system is initialized with random atomic displacements and equilibrated for 50 ps. The atomic velocities and positions are collected every 20 time steps in the following 1 ns.
Using the Wiener-Khintchine theorem \cite{Dove1993}, we can compute the power spectra $Z_q(\omega)$ at a given wave vector $q$ given the trajectory:
\begin{equation}
Z_q(\omega)=\frac{1}{2\pi}\sum_{\alpha,s}\frac{m_\alpha}{N_q} \left| \int dt e^{-i\omega t}  \sum_{i} \upsilon ^{(\alpha,s)}_{i}(t) e^{-iq\cdot r_{i}(t)}   \right|^{2} \label{eq.wiener}
\end{equation}
where $\upsilon$ is the velocity, $r$ is the position, $\alpha$ labels the atom type (i.e. Pb or Te), $s$ labels the displacement vector (i.e. $x$, $y$, $z$),
and $N_q$ is the number of $q$-points. The power spectrum is equivalent to the imaginary part of the phonon
Green's function in the classical limit.

\begin{figure}%[b]
\includegraphics[width=0.4\textwidth]{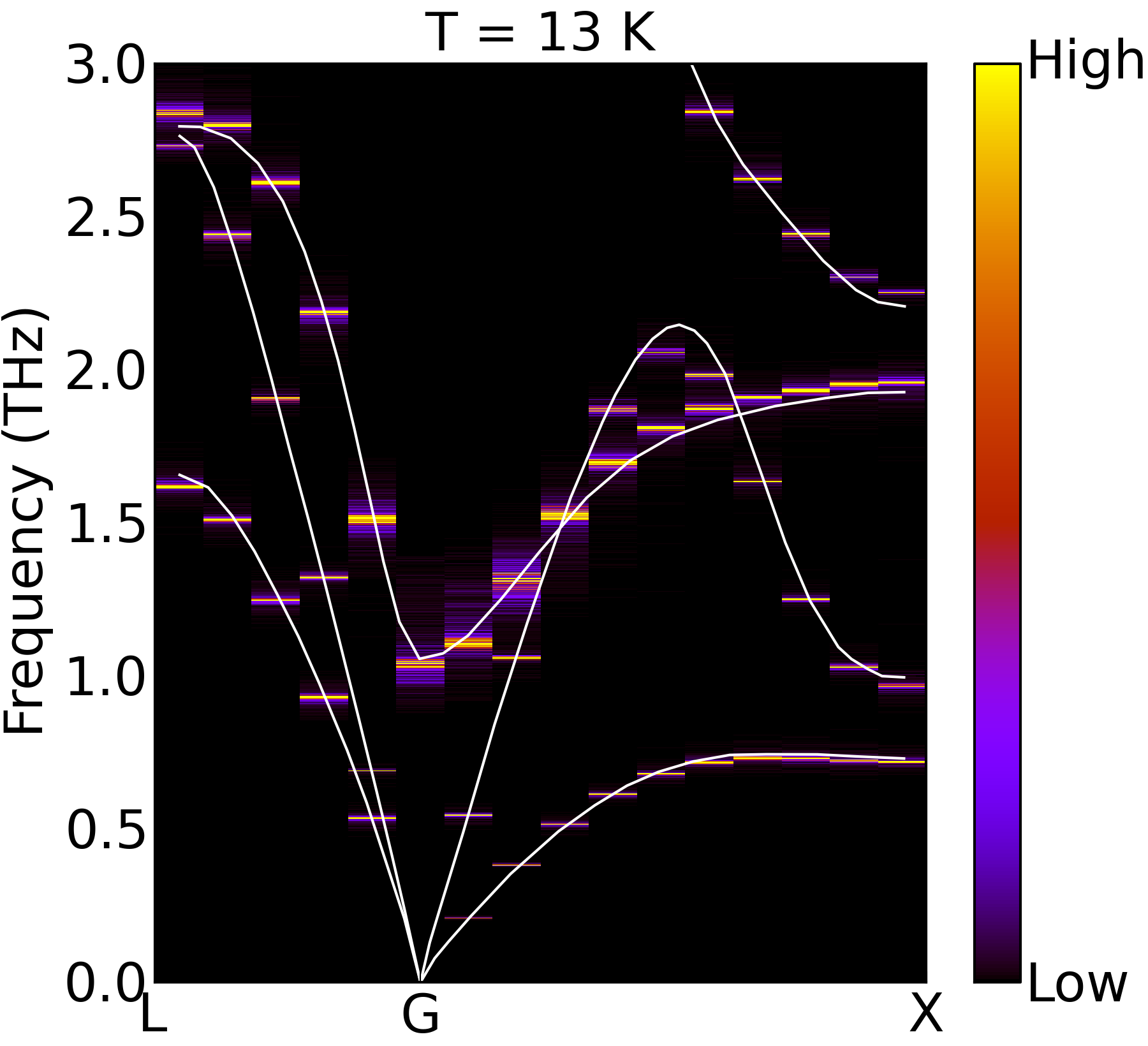} \\
\includegraphics[width=0.4\textwidth]{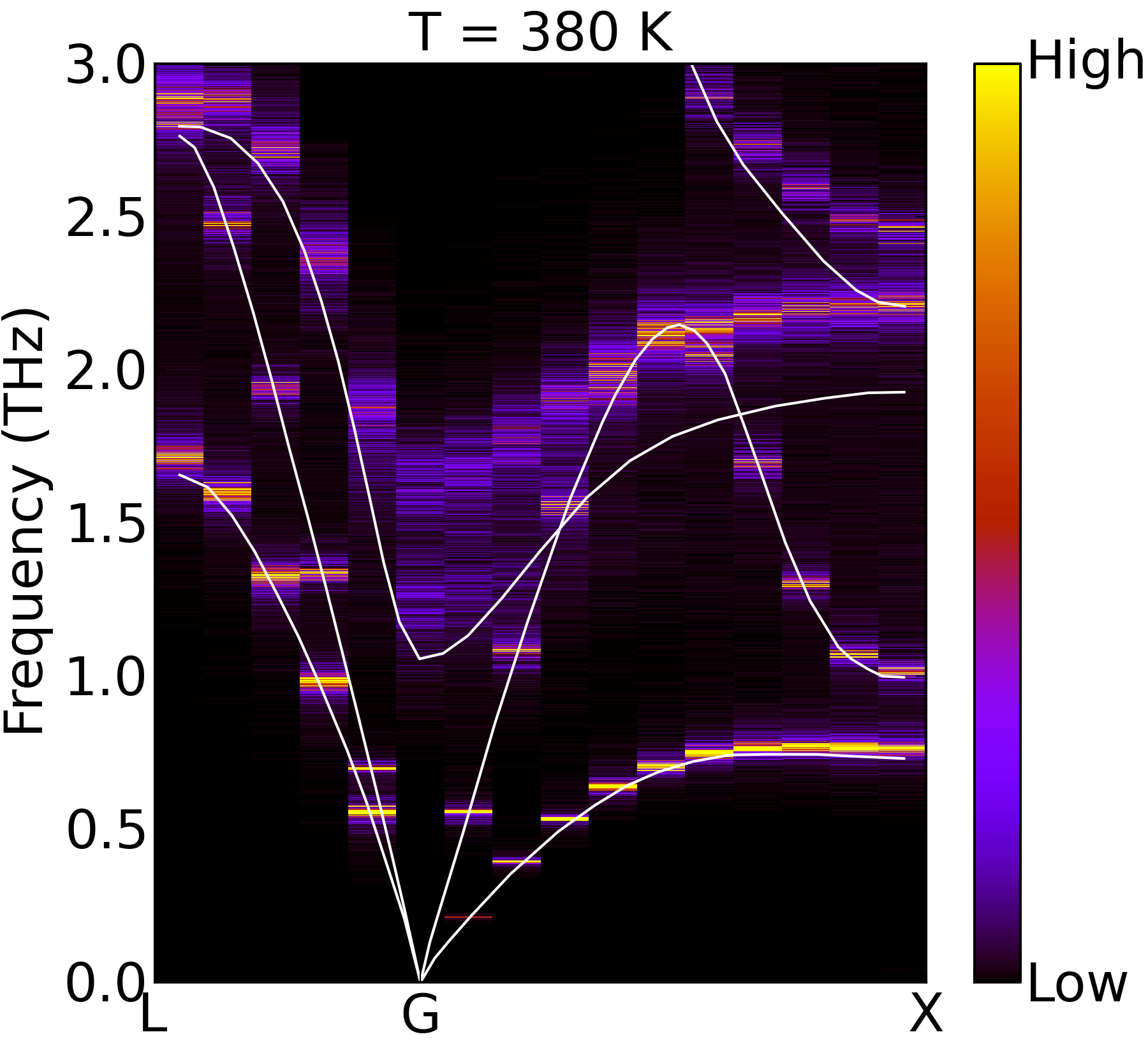}
\caption{\label{fig:phDisp}(Color online). The classical phonon spectra of PbTe along the
high symmetry directions in the first Brillouin zone. The white lines show the
0 K phonon dispersions calculated with the finite displacement method by Zhang
\emph{et al.} \cite{Zhang2009}.}
\end{figure}

The finite supercell size of 4000 primitive units in our MD simulations yields 4000 wavevectors
as a discrete grid over the Brillouin zone.
Fig.~\ref{fig:phDisp} shows the  spectra
along selected high symmetry lines in the first Brillouin zone.
At finite temperatures, the phonons will interact and no longer exist at a well defined frequency.
The most elementary many-body renormalizations are broadening (i.e. diminished lifetime)
and shifting of the peaks, both of which can be seen within perturbation theory \cite{Dove1993}.
The white curves
are the phonon dispersions (i.e. zero temperature) of Zhang \emph{et al.} \cite{Zhang2009},
and only mild renormalizations are observed at the relatively low temperature of $T=13$ K.
However, it is already evident that the acoustic modes are much less affected by the interactions as compared to
the optical modes.

At $T=380$ K, much more pronounced  effects are observed. The low energy
acoustic modes still show relatively mild broadening and shifts, while other
branches have substantially shifted and broadened.  The most interesting
behavior is observed at the zone center where the spectrum has remarkably split
into two broad peaks, as measured in experiment \cite{Delaire2011,Jensen2012}.
Additionally, the whole branch of TO phonons shift significantly to higher
frequencies at this temperature, which explains the striking experimental
observation of avoided crossing between the TO and LA branches in the $\Gamma\rightarrow X$
direction.

\begin{figure}%[b]
\includegraphics[width=0.23\textwidth]{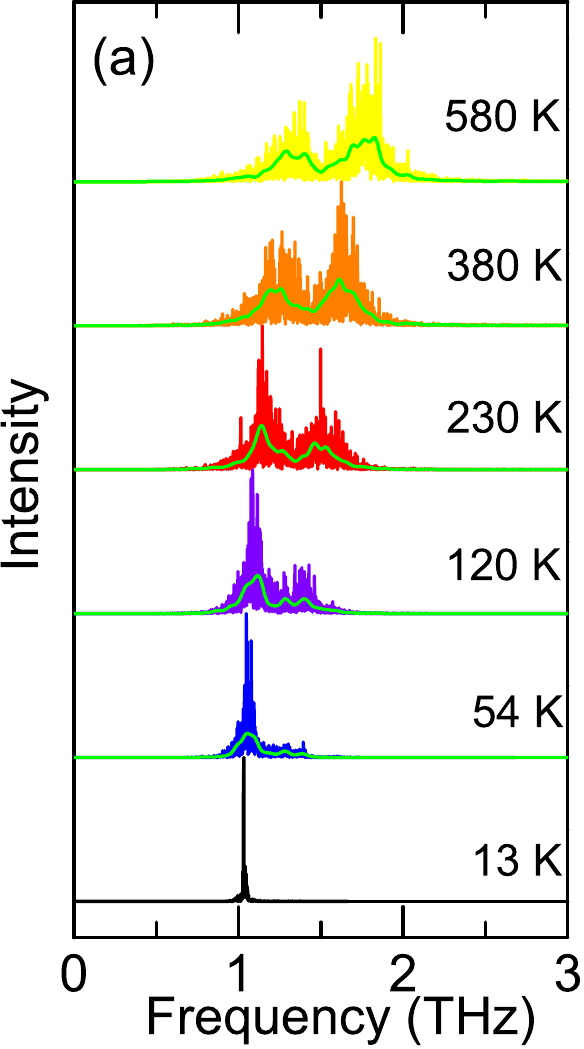}
\includegraphics[width=0.228\textwidth]{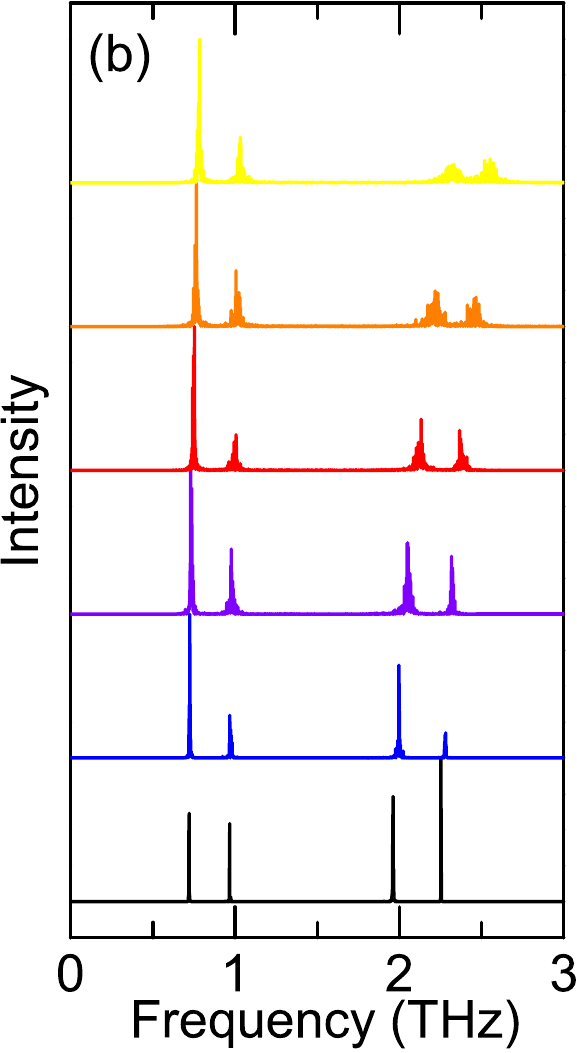}
\caption{\label{fig:phGX}(Color online). The classical phonon spectra at the zone center
(a) and the zone boundary $X$ (b) at different temperatures. The green curves
represent the data convolved with a Gaussian function of width 0.02 THz.}
\end{figure}

In order to further elucidate the behavior at the zone center,
the $q=0$ spectrum is computed  at a range of
temperatures (see Fig.~\ref{fig:phGX}a).
At the lowest temperature $T=13$ K the peak is sharp, while signatures
of nonlinearities can clearly be seen at $T=54$ K with the formation
a second feature at slightly higher energy. As the temperature is increased,
the center of gravity of the two peaks increases and spectral weight is transferred
from the lower peak to the higher peak.
Between $T=230$ K and $T=380$ K the splitting of the peaks changes relatively little, with a value  of $\approx 0.37$ THz (i.e. $\approx 1.5$ meV).
This is in good quantitative agreement with the measurements of Jensen \emph{et al.} \cite{Jensen2012} (i.e. $\approx 1.6$ meV) and
Delaire \emph{et al.} \cite{Delaire2011} (i.e. $\approx 1.9$ meV).
This behavior should be contrasted to the $X$-point, which behaves in a more simplistic fashion with mild
broadening and shifts of the peaks (see Fig.~\ref{fig:phGX}b).

\begin{figure}%[b]
\includegraphics[width=0.38\textwidth]{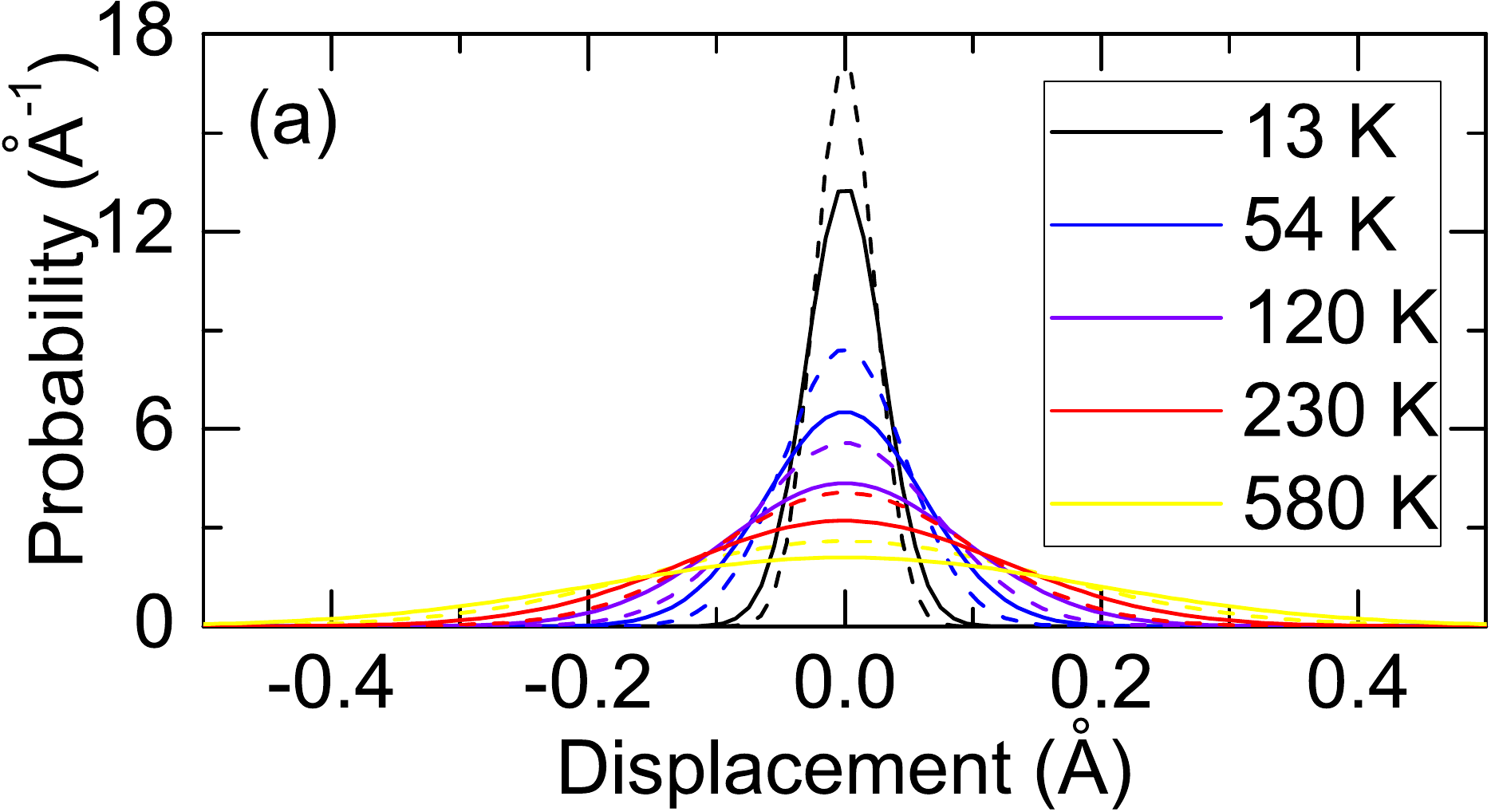} \\
\includegraphics[width=0.38\textwidth]{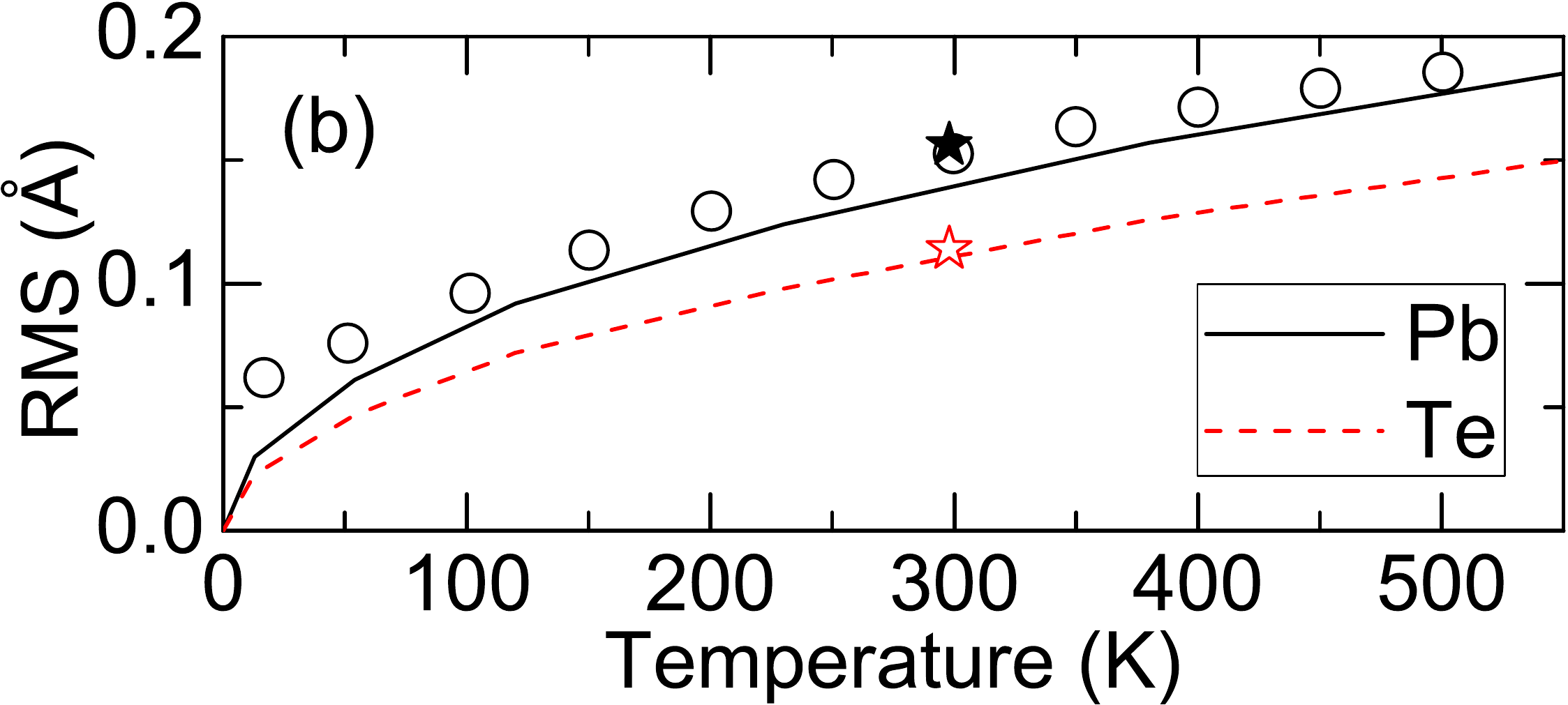}
\includegraphics[width=0.4\textwidth]{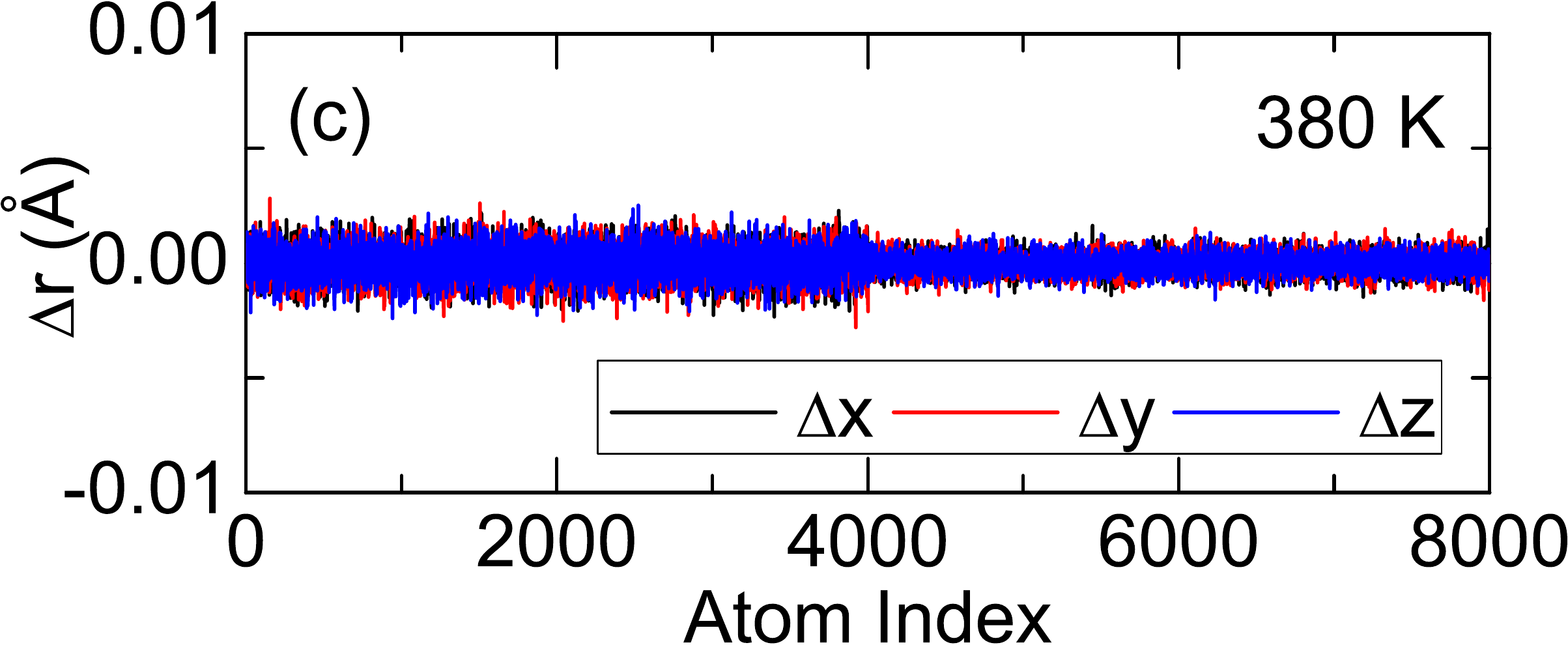}
\caption{\label{fig:probability}(Color online). (a) The reduced probability density for Pb (solid lines) and Te (dashed lines) at different temperatures. (b) The RMS displacement as a function of temperature. The square root of experimental atomic displacement parameters are shown as dots \cite{Bozin2010} and stars \cite{Pereira20131300}. (c) The average
displacements of each atoms in the supercell. The first 4000 atoms are Pb, followed by Te.}
\end{figure}

Jensen \emph{et al.} \cite{Jensen2012} proposed that the emergence of the new
spectral peak is related to a local symmetry breaking in PbTe at elevated
temperatures based on analysis of INS measurements. This local symmetry
breaking was first proposed in an earlier work where Pb was suggested to
achieve a large average local displacement, nearly 0.2 {\AA} at room
temperature, with a possible interpretation being that Pb fluctuates between symmetry equivalent sites \cite{Bozin2010}. Kastbjerg \emph{et al.} claim an
even larger off-center displacement of Pb based on synchrotron powder x-ray
diffraction \cite{Kastbjerg2013}.  However, x-ray absorption fine structure
spectroscopy experiments have been interpreted as showing no such symmetry
breaking \cite{Keiber2013}. Given that our method faithfully reproduces the temperature dependence of the spectra, we can clearly answer this question with the position probability distribution function $P_\alpha(x,y,z)$, where $\alpha$ labels a given Pb or Te atom. We compute the reduced probability density $P_r(x_1)=\frac{1}{3N_q}\sum_\alpha \int dx_2dx_3 (1+p_{12}+p_{13})P_\alpha(x_1,x_2,x_3)$, where $p_{ij}$ is the permutation operator. $P_r$ is symmetric and shows no indication of either ion dwelling away from zero displacement (Fig.~\ref{fig:probability}a). The average displacements (Fig.~\ref{fig:probability}c) of each ion is very small (i.e. less than 0.003 {\AA} of static displacement at 380 K). Furthermore, individually examining $P_\alpha$ shows that there is no appreciable local symmetry breaking. However, there is a large root mean square (RMS) displacement which compares favorably with experiment  (Fig.~\ref{fig:probability}b). It should also be noted that the asymmetry observed in the measured pair distribution function \cite{Bozin2010} and later reproduced in
AIMD \cite{Zhang2011} are successfully captured in our simulations.

\begin{figure}%[b]
\includegraphics[width=0.32\textwidth]{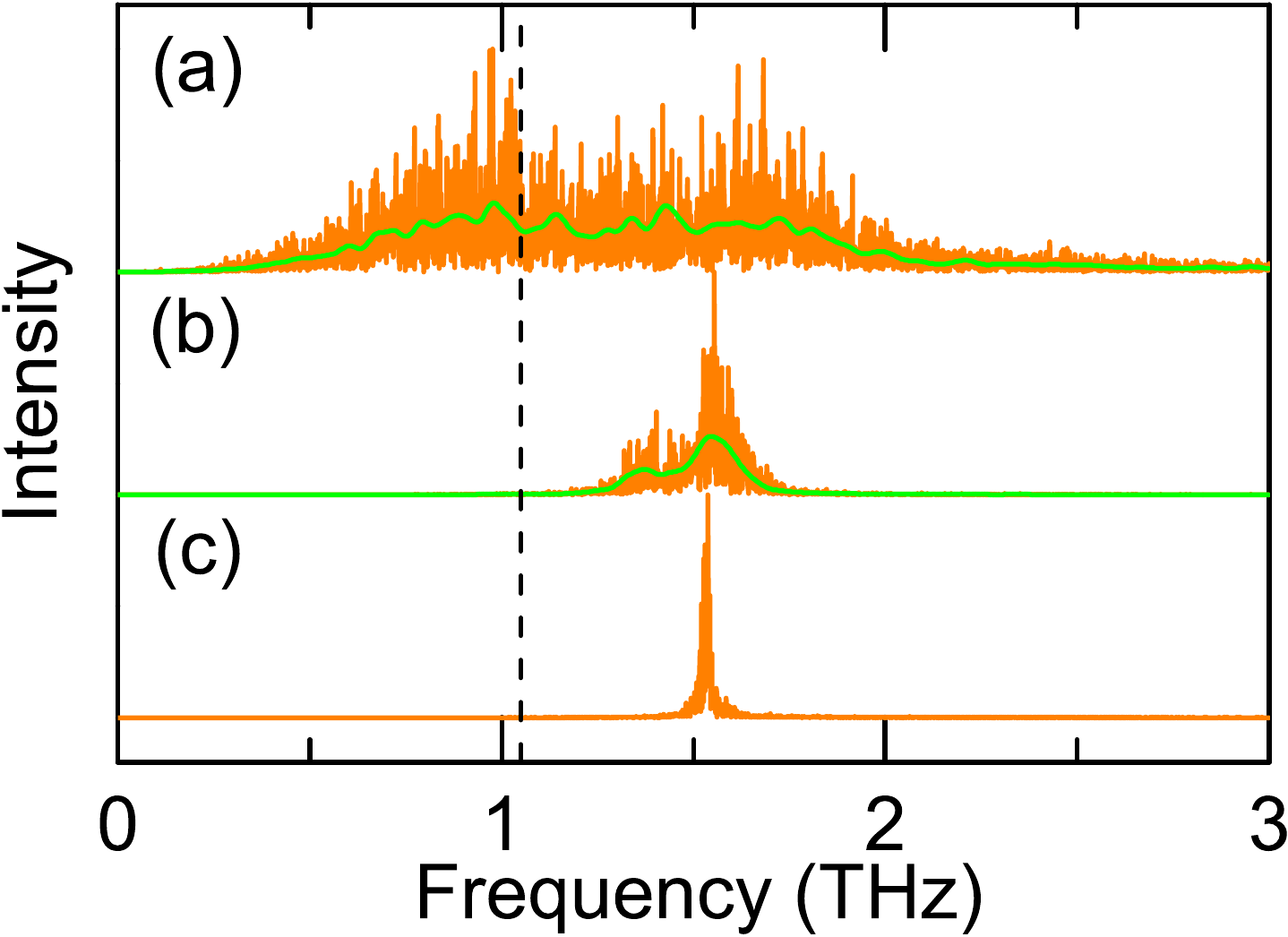}
\caption{\label{fig:Anh3}(Color online). The zone center classical phonon spectra at $T=380$ K
computed with fictitious values of $\Phi_3$ which are multiplied by a factor of
(a) $1.5$, (b) $0.5$ , and (c) $0$.
The green curves represent the data convolved with a Gaussian function of width 0.02 THz. The vertical dashed line is the $T=0$ K TO mode frequency.}
\end{figure}

This novel spitting of the TO peak arises from a broadened pole-like feature in
the self-energy which is generated either by the third order anharmonic term or
the fourth order term, and it is straightforward to separate the two.  We
cannot compute the spectra only using the third order term as the crystal
rapidly becomes unstable. However, the spectra can be computed with the third
order term set to zero, halved, and 1.5 times. When $\Phi_3=0$, the TO mode simply
shifts upward with very mild broadening (see Fig.~\ref{fig:Anh3}). Interestingly, the peak is located close to the upper peak in Fig.~\ref{fig:phGX}a. In the case where the
third order term is halved, the upper peak remains in a similar position but
the lower peak is now closer. Finally, when the third order term is amplified $1.5$ times,
there is massive scattering sending large amounts of spectral weight to low
energy. Therefore, we can see that the double peak is basically a confluence of
both terms, and the phenomenon is rather special in that it depends on a delicate balance
of two different terms.

In summary, we have demonstrated that using our slave mode expansion to compute
the classical phonon spectrum agrees remarkably well with INS measurements. The
anomalous splitting of the spectrum at the zone center is captured with our
approach. We show that there is no appreciable spontaneously broken symmetries,
resolving discrepancies in the experimental literature. All of this behavior
can be captured with only two anharmonic parameters, which are in essence
nearest neighbor cubic and quartic terms. Given that our work is classical, it
is clear that quantum fluctuations of the phonons appear to be irrelevant to
generating this spectral anomaly.  Future work will be important to
understanding this phenomenon in terms of traditional many-body techniques in the
semi-classical limit. More broadly, our approach should be applicable to materials across the
periodic table, and quantum mechanics can be incorporated with existing methods.

YC and CAM acknowledge  funding from a Columbia RISE grant. XA acknowledges support
from National Science Foundation (Grant No. CMMI-1150795). We thank
O. Delaire for stimulating conversations.

\bibliography{References}

\end{document}